\documentclass[a4paper,aps,prd,10pt,preprintnumbers,showpacs,twocolumn,superscriptaddress,nofootinbib,amsmath,amssymb]{revtex4-1}
\usepackage[dvips]{graphics}
\usepackage[utf8]{inputenc}
\usepackage[T1]{fontenc}
\usepackage{cmap}

\begin{document}
\title{Shadow of a black hole surrounded by dark matter}
\author{R. A. Konoplya}\email{roman.konoplya@gmail.com}
\affiliation{Institute of Physics and Research Centre of Theoretical Physics and Astrophysics, Faculty of Philosophy and Science, Silesian University in Opava, CZ-746 01 Opava, Czech Republic}
\affiliation{Peoples Friendship University of Russia (RUDN University), 6 Miklukho-Maklaya Street, Moscow 117198, Russian Federation}
\begin{abstract}
We consider a simple spherical model consisting of a Schwarzschild black hole of mass $M$ and a dark matter of mass $\Delta M$ around it. The general formula for the radius of black-hole shadow has been derived in this case. It is shown that the change of the shadow is not negligible, once  the effective radius of the dark matter halo is of order $\sim \sqrt{3 M \Delta M}$. For this to happen, for example, for the galactic black hole, the dark matter must be concentrated near the black hole. For small deviations from the Schwarzschild limit, the dominant contribution into the size  of a shadow is due to the dark matter under the photon sphere, but at larger deviations, the matter outside the photon sphere cannot be ignored.
\end{abstract}
\pacs{04.50.Kd,04.70.Bw,04.30.-w,04.80.Cc}
\maketitle

\section{Introduction}

Recent observations of black holes in the electromagnetic spectrum succeeded in observing the first image of the black hole in the center of galaxy M87 \cite{Akiyama:2019cqa,Akiyama:2019bqs}. Although the first image of the black hole does not allow one to identify the black hole geometry clearly, the principal strategy for the improvement of measurements should lead to much higher resolution in the future \cite{Goddi:2017pfy}. Therefore it is difficult to underestimate theoretical efforts to calculate  forms of shadows cast by black holes and black-hole mimickers in various theories of gravity and astrophysical environments \cite{Cunha:2016bjh}-\cite{Wei:2019pjf}.

At the same time, it is believed that $85 \%$ of mass in the Universe consists of the invisible dark matter \cite{Jarosik}. The abnormally high velocities of stars at the outskirts of galaxies imply that visible disks of galaxies are  immersed in a much larger roughly spherical halo of dark matter \cite{Kafle:2014xfa,Battaglia:2005rj}.  The dark matter does not interact with the electromagnetic field and therefore the propagation of light is possible inside the dark matter halo. The natural question would be to learn whether the black hole shadow could be affected by the tidal forces induced by the invisible matter. A few attempts in this direction have been made in \cite{Hou:2018avu,Haroon:2018ryd,Hou:2018bar,Cunha:2015yba} and a similar work was done for the dark energy in \cite{Abdujabbarov:2015pqp}. However, in all of the above works one or the other particular equation of state for the dark matter or dark energy was assumed, so that the results look highly model-dependent. For example, in the case of the dark energy \cite{Abdujabbarov:2015pqp}, one particular solution from the family of solutions obtained by Kiselev \cite{Kiselev:2002dx} was analyzed.

In our opinion, before considering particular models for the invisible matter, a simpler question must be answered: Can dark matter deform the black hole geometry so strongly, that the shadow would change seemingly? In order to answer this question we should imply only basic features of the dark matter: that it has a kind of an effective mass and does not interacting with the electromagnetic field, so that its influence on the black-hole shadow is only through changing the background geometry. For this purpose we will consider the spherically symmetric configuration, consisting of the Schwarzschild black hole and a spherical halo of dark matter  around it. This model was considered for testing the gravitational response (such as quasinormal modes \cite{Konoplya:2011qq} or echoes \cite{Cardoso:2017cqb}) of black holes in the astrophysical environment \cite{Leung:1999rh,Barausse:2014tra,Konoplya:2018yrp}.

Here we will use the above framework for estimating the effect of dark matter on the size of the shadow of a black hole.  The paper is organized as follows. Sec. II introduces the mass function and the essential properties of the space-time under consideration. Sec. III briefly relates the deduction of the formula for a shadow of an arbitrary spherically symmetric background. In Sec. IV we derive the main results of the paper, devoted to the general formula for the radius of the shadow in the presence of dark matter. Finally, in Sec. V we summarize the obtained results and discuss the open questions.

\section{Modelling dark matter}

There are various approaches to modelling dark matter in General Relativity, taking into account current cosmological observations.  Here we will try to explore a more agnostic approach and will use the two facts:

\begin{itemize}
\item The dark matter is invisible matter which does not interact with an electromagnetic field. This way it should allow for propagation of light rays. This statement may be not true for some models of barionic dark matter which may absorb light on its way to the observer. If one suppose that the heavy barionic dark matter absorbs light intensively, any conclusion depends on a particular distribution and equation of state for the dark matter.
\item The dark matter possesses some mass which can be modelled as an additional effective mass in the mass function of a black hole.
\end{itemize}

Therefore, we choose the metric for the above configuration in the following way
\begin{equation}\label{metric}
ds^2=-f(r)dt^2+\frac{dr^2}{f(r)}+r^2(d\theta^2+\sin^2\theta d\phi^2)\,,
\end{equation}
where $$f(r)=1-\frac{2m(r)}{r}\,.$$
and the mass function is given by
\begin{equation}\label{massfunc}
m(r)=\left\{
        \begin{array}{l}
          M,\phantom{\Delta} \qquad\qquad\qquad r<r_s\,; \\
          M+\Delta M\left(3-2\dfrac{r-r_s}{\Delta r_s}\right)\left(\dfrac{r-r_s}{\Delta r_s}\right)^2,  \\
          \phantom{\Delta~M,\!} \qquad\qquad\qquad r_s\leq r\leq  r_s+\Delta r_s\,;\\
          M + \Delta M, \qquad\qquad\qquad r_s+\Delta r_s<r\,.
        \end{array}
      \right.
\end{equation}

\begin{figure}
\resizebox{\linewidth}{!}{\includegraphics*{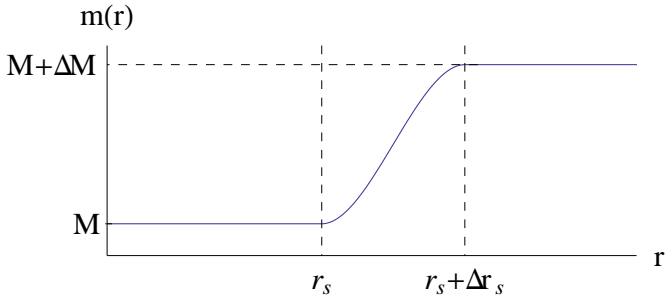}}
\caption{Choice of the mass function.}\label{fig:massfunc}
\end{figure}

This way $m(r)$ and $m'(r)$ are continuous functions (see Fig.~\ref{fig:massfunc}). Here $\Delta M > 0$ ($\Delta M < 0$) corresponds to positive (negative) energy density of matter. Although our primary motivation is to study positive $\Delta M$ (which corresponds to dark matter), for completeness we will also consider effect of negative mass, having in mind, as a by-product, possible exotic matter with a negative kinetic term or repulsion.

\begin{figure}
\resizebox{\linewidth}{!}{\includegraphics*{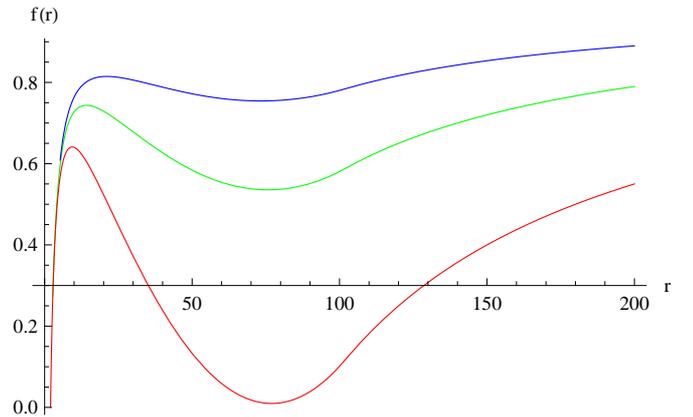}}
\caption{The metric function at $r_s = 2$, $\Delta r_s =100$, $\Delta M/M = 10$ (top), $20$, $44$ (bottom); $M=1$.}\label{fig:metricfun}
\end{figure}

Thus, the dark matter is situated in the region which begins from $r= r_s \geq 2 M$ and finishes at $r= r_s + \Delta r_s$, while the event horizon is still located at $r = 2 M$.
One should have in mind that for $\Delta M > 0$ and some fixed values of $r_s$ and $\Delta r_s$, the allowed values of $\Delta M$ are constrained, because if $\Delta M$ is too large, it simply increases the total mass of the black hole and the radius of the event horizon. As an example, from fig. \ref{fig:metricfun} one can see that when $\Delta M/M$ is increasing,  the bottom metric is approaching the state where the radius of the event horizon becomes larger at $\delta M/M \approx 45$ for given $r_s$ and $\Delta r_s$. The absolute value of the negative $\Delta M$ is not constrained in this way.

\section{General formula for the radius of shadow of spherically symmetric black holes}
Here, following \cite{Perlick:2015vta}, we will briefly present the deduction of the formula for a shadow of an arbitrary spherically symmetric black hole. The metric of a spherically symmetric spacetime can be written in the following way
\begin{equation}\label{eq:g}
d s^2 = - A(r) dt^2 + B(r) dr^2 +
D(r) \big( d \vartheta ^2 + \mathrm{sin} ^2 \vartheta \,
d \varphi ^2 \big) \, ,
\end{equation}
In spherical case, each plane can be considered as an equatorial one, so that we can choose $\vartheta = \pi /2$, and consequently $p_{\vartheta}=0$.
The Hamiltonian for light rays has the form
\begin{equation}\label{eq:H}
H  = \frac{1}{2}  g^{ik} p_{i} p_{k} = \frac{1}{2} \left( - \dfrac{p_t^2}{A(r)} + \dfrac{p_r^2}{B(r)}
+ \dfrac{p_{\varphi}^2}{D(r)} \right).
\end{equation}
The light rays are the solutions to the equations of motion:
\begin{equation}\label{eq:Ham0}
\dot{p}{}_i = -\dfrac{\partial H}{\partial x^i}
\, , \quad
\dot{x}{}^i = \dfrac{\partial H}{\partial p_i},
\end{equation}
so that
\begin{equation}\label{eq:Ham4}
\dot{t} \,  = \, - \, \dfrac{p_t}{A(r)}   \, ,
\end{equation}
\begin{equation}\label{eq:Ham5}
\dot{\varphi} \, = \,
\dfrac{p_{\varphi}}{D(r)}   \, ,
\end{equation}
\begin{equation}\label{eq:Ham6}
\dot{r} \,  = \,
\dfrac{p_r}{B(r)}.
\end{equation}
From $H=0$ it follows that
\begin{equation}\label{eq:H0}
0 \, = \, - \dfrac{p_t^2}{A(r)} + \dfrac{p_r^2}{B(r)} + \dfrac{p_{\varphi}^2}{D(r)}.
\end{equation}
Here a dot designates derivatives with respect to an affine parameter and a prime is for derivatives with respect to $r$.
The momenta $p_t$ and $p_\varphi$ are constants of motion; $\omega _0 := - p_t$.

From (\ref{eq:Ham5}) and (\ref{eq:Ham6}) one finds that
\begin{equation}\label{eq:drdphi1}
\dfrac{dr}{d \varphi}  \, = \,
\dfrac{\dot{r}}{\dot{\varphi}}
\, = \,
\dfrac{D(r) p_r}{B(r) p_{\varphi}}   \, .
\end{equation}
Using $p_r$ from (\ref{eq:H0}), we have
\begin{equation}\label{eq:drdphi2}
\dfrac{dr}{d \varphi}  \, = \,
\pm \, \dfrac{\sqrt{D(r)} }{\sqrt{B(r)}}
\sqrt{\dfrac{\omega _0^2}{p_{\varphi} ^2} \, h(r)^2
\, - \, 1 \,}
\end{equation}
where, following \cite{Perlick:2015vta}, the function $h(r)$ is defined as follows:
\begin{equation} \label{eq:h-definition}
h(r)^2 = \dfrac{D(r)}{A(r)} .
\end{equation}
A circular light orbit corresponds to zero radial velocity and acceleration, so that  $\dot{r}=0$ and $\ddot{r}=0$. From (\ref{eq:Ham6})
it follows that $p_r=0$, while from (\ref{eq:H0}) we find
\begin{equation}\label{eq:circ1}
0 \, = \, - \dfrac{\omega _0^2}{A(r)}
+ \dfrac{p_{\varphi}^2}{D(r)}.
\end{equation}
Differentiating (\ref{eq:Ham6}) with respect to the affine parameter gives
\begin{equation}
\dot{p}_r = \frac{d}{d \lambda}
\big( B(r) \, \dot{r} \big) =
\ddot{r} B(r) + \dot{r}{}^2 B'(r) \, .
\end{equation}
Then, the requirement of zero radial velocity and acceleration
leads to $\dot{p}_r=0$, and we find that
\begin{equation}\label{eq:circ2}
0 \, = \, - \,
\dfrac{\omega _0^2A'(r)}{A(r)^2}
\, + \, \dfrac{p_{\varphi}^2D'(r)}{D(r)^2}.
\end{equation}
From (\ref{eq:circ1}) and (\ref{eq:circ2}) it follows that
\begin{equation}\label{eq:circ3}
p_{\varphi}^2 \, = \, D(r) \Big(
\dfrac{\omega _0^2}{A(r)} \Big) =\dfrac{D(r)^2}{D'(r)} \Big( \dfrac{\omega _0^2 A'(r)}{A(r)^2} \Big).
\end{equation}
Then, the radius of a photon sphere is a solution to the equation
\begin{equation}\label{eq:circ8}
0 \, = \, \dfrac{d}{dr} h(r)^2.
\end{equation}
Following the designations of \cite{Perlick:2015vta}, we will use $r_{\mathrm{O}}$ for the position of the observer and $\alpha$ for the angle respectively the radial direction.
Then, we have
\begin{equation}\label{eq:alpha1}
\cot \, \alpha \, =
\left. \frac{\sqrt{g_{rr}}}{\sqrt{g_{\varphi \varphi}}}
\, \dfrac{dr}{d \varphi} \right|_{r=r_{\mathrm{O}}} = \left.
\dfrac{\sqrt{B(r)}}{\sqrt{D(r)}} \,
\dfrac{dr}{d \varphi} \right|_{r=r_{\mathrm{O}}} \, .
\end{equation}
The equation (\ref{eq:drdphi2}) can be
rewritten in terms of the minimal radius $R$ as follows
\begin{equation}
\frac{dr}{d \varphi} =
\pm \dfrac{\sqrt{D(r)}}{\sqrt{B(r)}}
\sqrt{\dfrac{h^2(r)}{h^2(R)} \, - \, 1 \,} \, .
\end{equation}
Then we have
\begin{equation}
\cot^2 \alpha \, = \frac{h^2(r_{\mathrm{O}})}{h^2(R)}  - 1 \, .
\end{equation}
and consequently
\begin{equation}\label{eq:sinalpha}
\mathrm{sin} ^2 \alpha \, = \, \dfrac{h(R)^2}{h(r_{\mathrm{O}})^2}   \, .
\end{equation}
The angular radius of the shadow is then determined by
\begin{equation}\label{eq:shadow}
\mathrm{sin} ^2 \alpha_{\mathrm{sh}} \, = \,
\dfrac{h(r_{\mathrm{ph}})^2}{h(r_{\mathrm{O}})^2}.
\end{equation}
With these formulas at hand we are ready to find the radius of the shadow for the black hole metric discussed in the previous section.

\section{Shadow of a black hole surrounded by dark matter}

\begin{figure}
\resizebox{\linewidth}{!}{\includegraphics*{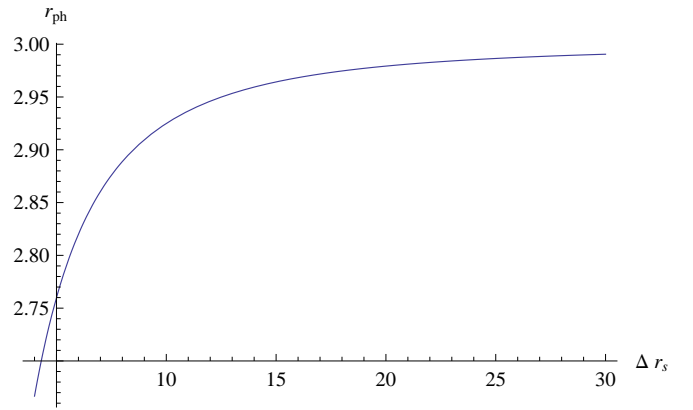}}
\caption{The radius of the closest to the black hole photon sphere as a function of $\Delta r_{s}$, $M=1$, $\Delta M = M$.}\label{fig:PhotonSphere0}
\end{figure}

\begin{figure}
\resizebox{\linewidth}{!}{\includegraphics*{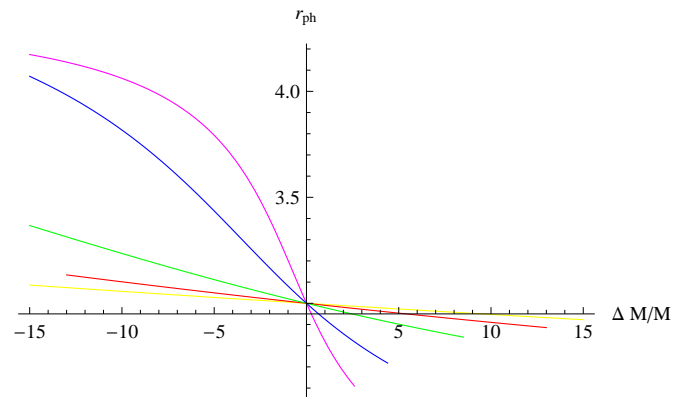}}
\caption{The radius of the closest to the black hole photon sphere as a function of $\Delta M$ for $\Delta r_s = 3$ (right - bottom), $5$, $10$, $15$, $20$ (right - top); $r_0 =1$, $r_s = 1.01$.}\label{fig:PhotonSphere1}
\end{figure}

\begin{figure}
\resizebox{\linewidth}{!}{\includegraphics*{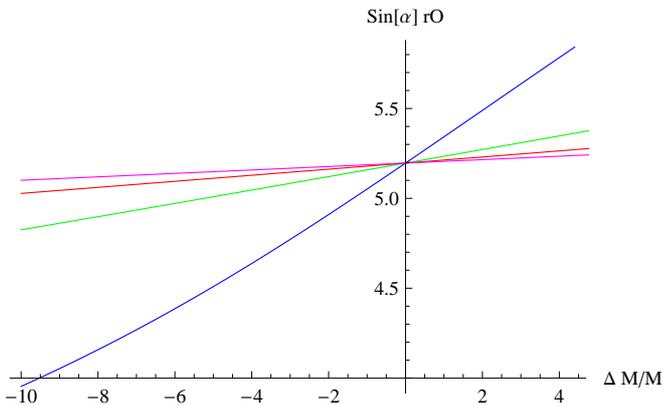}}
\caption{The radius of the black hole shadow as a function of $\Delta M/M$ for $\Delta r_s = 10$ (right - bottom), $20$, $30$, $40$ (right - top); $r_s = r_{0}$.}\label{fig:PhotonSphere2}
\end{figure}

Depending on the position of the photon sphere relatively the dark matter halo, there are three qualitatively different situations:

\begin{enumerate}
\item The dark matter is distributed in such a way that the photon sphere lies between the even horizon $r_{0}$ and the beginning of the dark matter layer $r_{s}$,
$$
r_{0}\leq r_{ph} < r_s,
$$
where $r_0 = 2 M$ is the Schwarzschild radius.
Then for spherically symmetric configuration the observer will see the purely Schwarzschild shadow, corresponding to the mass $M$, while the photon sphere is simply
\begin{equation}
r_{ph} = 3 M.
\end{equation}

\item The photon sphere lies outside the dark matter configuration on the side of the observer, that is between the observer and the beginning of the dark matter,
$$
r_{s}+ \Delta r_s < r_{ph} < r_{\mathrm{O}}.
$$
Then the observed photon sphere has simply an added mass of the dark matter:
\begin{equation}
r_{ph} = 3 (M + \Delta M).
\end{equation}
This situation does not look realistic as then it would mean that the dark matter exists only in the proximity of the black hole and, at the same time, does not fall onto the black hole.

\item The only non-trivial situation occurs when the closest to the horizon photon orbit is placed inside the dark matter configuration, so that
$$
r_{s} < r_{ph} < r_{s}+ \Delta r_s.
$$
In that case, solution of the equation (\ref{eq:circ8}) for the metric functions defined in (\ref{metric},\ref{massfunc}) give the following expression for the radius of the closest photon sphere:
\end{enumerate}
\begin{equation}\label{rph}
r_{ph} =  \frac{-\sqrt{K} \text{$\Delta $r}_s+6 \text{$\Delta $M} r_s \left(r_s+\text{$\Delta $r}_s\right)+\text{$\Delta $r}_s^3}{3
   \text{$\Delta $M} \left(2 r_s+\text{$\Delta $r}_s\right)},
\end{equation}
where
\begin{eqnarray}\label{eq:scheme}
K&=&  12 \text{$\Delta $M} r_s \text{$\Delta $r}_s \left(\text{$\Delta $r}_s-3 M\right)  \\
&  &  -18 M \text{$\Delta $M} \text{$\Delta $r}_s^2+3
   \text{$\Delta $M} r_s^2 \left(4 \text{$\Delta $r}_s+3 \text{$\Delta $M}\right)+\text{$\Delta $r}_s^4.
\nonumber
\end{eqnarray}
There are two photon spheres: the other one is given by the same equation  (\ref{rph}), where $K$ comes with an opposite sign in eq. (\ref{rph}). It is situated far from the black hole and is irrelevant for our consideration.

The function $h(r)$ is given by
\begin{equation}
h(r)^2= \frac{r^2}{1-\frac{2 M \left(\frac{\text{$\Delta $M} \left(r-r_s\right){}^2 \left(2 r_s-2 r+3 \text{$\Delta $r}_s\right)}{2 M
   \text{$\Delta $r}_s^3}+1\right)}{r}}.
\end{equation}
The matter under the photon sphere should then move towards the horizon and a more realistic model should include non-static configuration of matter. However, here we are interested in a robust description of \emph{the order} of deviation of the black-hole shadow owing to the effective mass of the black-hole environment.The estimation of orders of the effect should give us the first understanding whether such a global astrophysical factor as the dark matter can influence the black hole shadow at all.

From fig. (\ref{fig:PhotonSphere0}) one can see that once the mass of the dark matter is fixed, and the distance $\Delta r_s$ over which this mass is distributed is increased, then the diminished density of the dark matter leads to larger values of the radius of the photon sphere. In the limit of zero density this radius approaches its Schwarzschild value $r= 3 M$. Although the radius of the photon sphere decreases when the mass $\Delta M$ is increasing (see fig. (\ref{fig:PhotonSphere1})), from fig. (\ref{fig:PhotonSphere2}), one can notice that the effect for the shadow is opposite: larger masses of dark matter correspond to larger radii of shadows. From the first sight it looks counterintuitive. From the naive, purely Newtonian analogy, the radius of the photon sphere should be determined by the total mass of matter inside it. Then, the larger $\Delta M$, the larger must be an equilibrium orbit compensating the gravitational attraction. In reality this logic is not additive, because the matter is situated not only under the photon sphere, but also outside it, so that this increased orbit would imply that even larger mass is situated under the new ``photon sphere'', which would require even larger radius of the photon orbit, and so on. Therefore, when $\Delta M$ is increased, \emph{a smaller} radius of the photon sphere corresponds to a new position of equilibrium. The radius of the shadow depends on the propagation of light in the whole space between an observer and a black hole and, therefore, may differ from the photon sphere behavior, which we observe in the considered model.

The analytical expression for the radius of shadow as a function of $M$, $\Delta M$, $r_s$ and $\delta r_s$ is rather cumbersome, but a concise  expression can be found for sufficiently large $\Delta r_s$, that is, for the astrophysically most expected situation in which the cloud of dark matter is distributed over the whole halo rather then concentrated in a single place. The Taylor expansion for large $\Delta r_s$ gives
\begin{eqnarray}
 \mathrm{sin} \alpha_{\mathrm{sh}}r_{\mathrm{O}} &=& 3 \sqrt{3} M+\frac{9 \sqrt{3}  \text{$\Delta $M} \left(3 M- r_s\right){}^2}{ \text{$\Delta $r}_s^2} \\
 \nonumber
& &  - \frac{3 \sqrt{3}
   \text{$\Delta $M} \left(3 M- r_s\right){}^3}{\text{$\Delta $r}_s^3} \\
   \nonumber
 & &  + \frac{162 \sqrt{3} \text{$\Delta $M}^2 \left(2 M-r_s\right) \left(r_s-3 M\right){}^2}{\text{$\Delta $r}_s^4} \\
 \nonumber
& & +O\left(\frac{M}{\text{$\Delta $r}_s}\right)^5.
\end{eqnarray}
If we suppose that the dark matter surrounds the black hole right from its event horizon, then, $r_s = 2 M$ and the above formula is reduced to the following form
$$ \mathrm{sin} \alpha_{\mathrm{sh}}r_{\mathrm{O}} \approx 3 \sqrt{3} M+\frac{9 \sqrt{3} M^2 \text{$\Delta $M}}{\text{$\Delta $r}_s^2}-\frac{6 \left(\sqrt{3} M^3 \text{$\Delta  $M}\right)}{\text{$\Delta $r}_s^3} $$
 \begin{equation}
+\frac{54 \sqrt{3} M^4 \text{$\Delta $M}^2}{\text{$\Delta
   $r}_s^5}+O\left(\frac{M}{\text{$\Delta $r}_s}\right){}^6.
 \end{equation}
That means that for the black-hole shadow to be considerably changed by the dark matter, one should have
\begin{equation}
3 \Delta M \cdot M \sim \Delta r_{s}^2,
\end{equation}
which is not fulfilled for the central black hole in our galaxy, because from estimations of the mass of dark matter halo in our galaxy, we know that  $\Delta M \approx 6 \cdot 10^{11}- 3 \cdot 10^{12} M_{Sun}$ \cite{Kafle:2014xfa}, while the mass of the galactic black hole is $M \approx 4.3 \cdot 10^{6} M_{Sun}$. This leads to values of $\Delta r_s$ which are many orders smaller than the characteristic sizes of galaxies or dark matter halos.

\begin{figure}
\resizebox{\linewidth}{!}{\includegraphics*{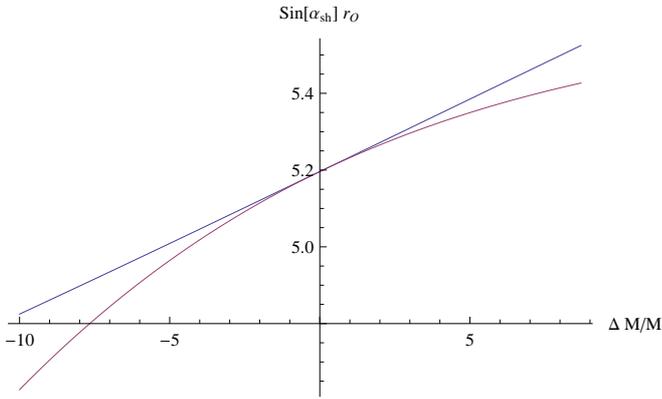}}
\caption{The radius of the black hole shadow as a function of $\Delta M/M$ calculated via differentiation of $h(r)^2$ (bottom) versus the one obtained from eq. (\ref{31}) (supposing that the matter outside the photon sphere is ignored); $\Delta r_s =20$, $r_s = 2 M$.}\label{fig:ShadowMe3}
\end{figure}

It is natural to suppose that one could use the fact that the gravitational force of a spherically symmetric distribution of matter, acting on an incremental element of the matter at any given radius $r$ is only due to the matter inside the radius $r$, while exterior matter can be neglected. Then, intuitively we could expect that the radius of the shadow for a given value of the photon sphere could simply be calculated from the supposition that the dark matter \emph{only inside} the photon sphere is important, while the dark matter outside the photon sphere can be completely ignored. Then, the Schwarzschild formula for the radius of the shadow $\mathrm{sin} \alpha_{\mathrm{sh}}r_{\mathrm{O}} = 3 \sqrt{3} M$ could simply be altered by adding the fraction of $\Delta M$ which lies inside the photon sphere.
In other words, using the expression for the mass function (\ref{massfunc}), the radius of the shadow would be
\begin{eqnarray}
\nonumber
\frac{\mathrm{sin} \alpha_{\mathrm{sh}}r_{\mathrm{O}}}{3 \sqrt{3}}   \approx  M +\text{$\Delta $M}\\
\label{31}
- \frac{\text{$\Delta $M} \left(-2 r_s+2 r_{ph}+\text{$\Delta $r}_s\right) \left(r_s-r_{ph}+\text{$\Delta
   $r}_s\right){}^2}{\text{$\Delta $r}_s^3}.
\end{eqnarray}
However, the roots given by this equations differs from the one obtained earlier via direct differentiation of $h(r)^2$, as can be seen on fig. (\ref{fig:ShadowMe3}). In this way we can see that the black-hole shadow depends not only on the matter under the photon sphere, but also on the of matter outside it. However, once $\Delta r_s$ is large, the Taylor expansion of the above expression produces exactly the same first three terms, but differs in higher order terms:
\begin{eqnarray*}
 \mathrm{sin} \alpha_{\mathrm{sh}}r_{\mathrm{O}} \approx 3 \sqrt{3} M+\frac{9 \sqrt{3} \text{$\Delta $M} \left(r_s-3 M\right){}^2}{\text{$\Delta $r}_s^2}- \\
 \nonumber
 \frac{6 \sqrt{3}
   \text{$\Delta $M} \left(3 M-r_s\right){}^3}{\text{$\Delta $r}_s^3}+\frac{162 \sqrt{3} \text{$\Delta $M}^2
   \left(M-r_s\right) \left(r_s-3 M\right){}^2}{\text{$\Delta $r}_s^4}\\
    \nonumber
-\frac{54 \sqrt{3} \text{$\Delta $M}^2 \left(3 M-5
   r_s\right) \left(3 M-r_s\right){}^3}{\text{$\Delta $r}_s^5}+O\left(\frac{M}{\text{$\Delta
   $r}_s}\right)^6.
\end{eqnarray*}
Therefore, the matter under the photon sphere describes  relatively small deviations from the Schwarzschild limit very well. Strong deviations, when $\Delta M/M$ is so large, that the metric function is close to the creating a new position for the event horizon (for example, as on fig. \ref{fig:metricfun} for $\Delta M/M \approx 44$), cannot be calculated from the formula (\ref{31}), as the relative error becomes of the order of the effect in that case. This confirms that once the deviation of the shadow from its Schwarzschild limit is not small, the matter outside the photon sphere cannot be ignored, when calculating the full effect.

\section{Discussions}

Here we found an analytical expression for the radius of the black hole shadow, supposing a simple spherical configuration of dark matter around it. A robust estimates show that the dark matter is unlikely to manifest itself in the shadows of galactic black holes, unless its concentration near the black hole is abnormally high. Furthermore, if one believes that such a high concentration of dark matter is possible, he could study more accurate models for the distribution of dark matter, include rotation of a black hole and dark matter and consider various equations of state.
High deviations from spherical distribution of dark matter in the halo (if confirmed) would apparently distort the shape of the shadow as well.
It is worthwhile noticing that when the deviation of the radius of shadow from its Schwarzschild value is small, the dominant contribution into the size of a shadow is due to the dark matter under the photon sphere. Nevertheless, this is not so for large deviations from Schwarzschild limit and the matter outside the photon sphere cannot be ignored in that case.

\acknowledgments{
The author acknowledges  the  support  of  the  grant  19-03950S of Czech Science Foundation ($GA\check{C}R$) and Alexander Zhidenko for useful discussions. This publication has been prepared with the support of the “RUDN University Program 5-100”.}

\end{document}